\documentclass[a4paper,11pt]{article}
\pdfoutput=1 

\usepackage[usenames,dvipsnames,svgnames,table]{xcolor} 
\usepackage{jcapmod}
\usepackage{verbatim}

\usepackage[T1]{fontenc} 
\usepackage[latin9]{inputenc}
\definecolor{lightgray}{gray}{0.9}
\definecolor{Amber}{rgb}{1.0, 0.75, 0.0}
\definecolor{blizzardblue}{rgb}{0.67, 0.9, 0.93}

\usepackage{float}
\usepackage{graphicx}
\usepackage{hyperref}
\usepackage{amsmath}
\usepackage{amssymb}
\usepackage{microtype}
\usepackage{amsbsy}
\usepackage{color}
\usepackage[capitalise]{cleveref}
\usepackage{breakurl}
\usepackage{csquotes}
\usepackage{lmodern}
\usepackage{bm}
\usepackage{dsfont}

\pdfpageheight\paperheight
\pdfpagewidth\paperwidth

\renewcommand*{\vec}[1]{\bm{#1}}
\newcommand*{\unitvec}[1]{\vec{\hat{#1}}}
\newcommand*{\mat}[1]{\bm{\mathsf{#1}}}
\DeclareMathOperator{\diag}{diag}
\newcommand{\identity}{\ensuremath{\mathds{1}}}

\newcommand*{\E}[1]{\texorpdfstring{\ensuremath{E_{#1}}}{E#1}}
\newcommand*{\A}[1]{\texorpdfstring{\ensuremath{A_{#1}}}{A#1}}
\newcommand*{\Espace}{\texorpdfstring{\ensuremath{E^3}}{E³}}

\allowdisplaybreaks

\makeatletter
\DeclareRobustCommand{\rcite}[1]{%
  \rcite@aux#1,\@nil{#1}%
}
\def\rcite@aux#1,#2\@nil#3{%
  \if\relax#2\relax
    Ref.~\cite{#3}%
  \else
    Refs.~\cite{#3}%
  \fi
}
\makeatother

\subheader{IFT-UAM/CSIC-22-137}

\title{Cosmic topology. Part I. Limits on orientable Euclidean manifolds from circle searches}

\author[a]{Pip Petersen,}
\author[a,b,c]{Yashar Akrami,}
\author[a]{Craig J. Copi,}
\author[c]{Andrew H. Jaffe,}
\author[d]{Arthur Kosowsky,}
\author[a]{Deyan P. Mihaylov,}
\author[a,c]{Glenn D. Starkman,}
\author[a]{Andrius Tamosiunas,}
\author[e,c]{Johannes R. Eskilt,}
\author[a]{\"{O}zen\c{c} G\"{u}ng\"{o}r,}
\author[a]{Samanta Saha,}
\author[a]{and Quinn Taylor}

\collaboration{(COMPACT Collaboration)}
\collaborationImg{\includegraphics[width = 0.1\textwidth]{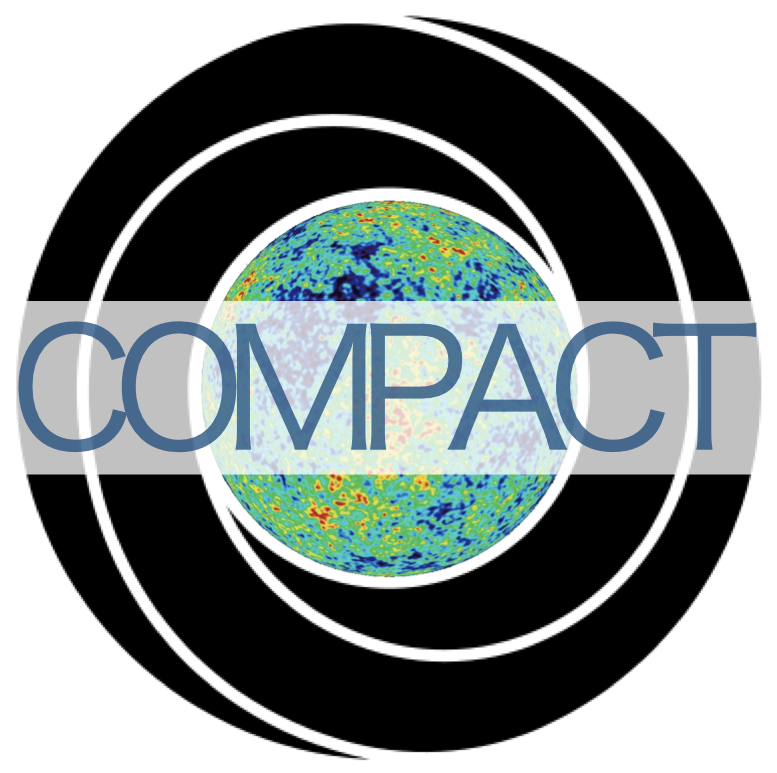}}

\affiliation[a]{CERCA/ISO, Department of Physics, Case Western Reserve University, 10900 Euclid Avenue, Cleveland, OH 44106, USA}
\affiliation[b]{Instituto de F\'isica Te\'orica (IFT) UAM-CSIC, C/ Nicol\'as Cabrera 13-15, Campus de Cantoblanco UAM, 28049 Madrid, Spain}
\affiliation[c]{Astrophysics Group \& Imperial Centre for Inference and Cosmology, Department of Physics, Imperial College London, Blackett Laboratory, Prince Consort Road, London SW7 2AZ, United Kingdom}
\affiliation[d]{Department of Physics and Astronomy, University of Pittsburgh, Pittsburgh, PA 15260, USA}
\affiliation[e]{Institute of Theoretical Astrophysics, University of Oslo, P.O. Box 1029 Blindern, N-0315 Oslo, Norway}

\emailAdd{petersenpip@case.edu}
\emailAdd{yashar.akrami@csic.es}
\emailAdd{craig.copi@case.edu}
\emailAdd{a.jaffe@imperial.ac.uk}
\emailAdd{kosowsky@pitt.edu}
\emailAdd{deyan.mihaylov@case.edu}
\emailAdd{glenn.starkman@case.edu}
\emailAdd{andrius.tamosiunas@case.edu}

\date{\today}

\abstract{The Einstein field equations of general relativity constrain the local curvature at every point in spacetime, but say nothing about the global topology of the Universe.
Cosmic microwave background anisotropies have proven to be the most powerful probe of non-trivial topology since, within $\Lambda$CDM, these anisotropies have well-characterized statistical properties, the signal is principally from a thin spherical shell centered on the observer (the last scattering surface), and space-based observations nearly cover the full sky.
The most generic signature of cosmic topology in the microwave background is pairs of circles with matching temperature and polarization patterns.
No such circle pairs have been seen above noise in the WMAP or \textit{Planck} temperature data, implying that the shortest non-contractible loop around the Universe through our location is longer than 98.5\% of the comoving diameter of the last scattering surface. 
We translate this generic constraint into limits on the parameters that characterize manifolds with each of the nine possible non-trivial orientable Euclidean topologies, and provide a code which computes these constraints.
In all but the simplest cases, the shortest non-contractible loop in the space can avoid us, and be shorter than the diameter of the last scattering surface by a factor ranging from 2 to at least 6.
This result implies that a broader range of manifolds is observationally allowed than widely appreciated.
Probing these manifolds will require more subtle statistical signatures than matched circles, such as off-diagonal correlations of harmonic coefficients.}

\keywords{cosmic topology, cosmic anomalies, statistical isotropy, cosmic microwave background, large-scale structure}

\begin{document}
\maketitle
\flushbottom

\section{Introduction}
\label{sec:intro}

The basic assumption of most cosmological models is that spacetime is isotropic and homogeneous as encoded in the Friedmann-Lema\^itre-Robertson-Walker metric. 
However, this metric is only an expression of the local spacetime geometry, and says nothing about the global topology.
Cosmologists have long speculated \cite{Lachieze-Rey:1995} about the possibility of a non-trivial spatial topology, which would contain non-contractible space-like loops: travelling far enough in a particular direction returns back to the original position.

Perhaps the most compelling case for our spatial manifold having a non-trivial topology is `why not?' 
We expect that a theory of quantum gravity may contain topology-changing processes, e.g., quantum foam \cite{HAWKING1978349, Carlip:2022pyh}.
If so, there would be no particular reason that, when the Universe emerges from the very early quantum-gravity era, the space would be the infinite-volume covering space of one of the eight possible homogeneous 3-geometries.\footnote{
    The covering space of a given geometry, such as $E^3$ (zero curvature), $S^3$ (positive curvature), or $H^3$ (negative curvature) is the manifold with no non-contractible closed loops, i.e., it is the manifold with trivial topology. It is also the manifold with the full isometry group of the geometry.}
Subsequent evolution of the Universe could hide the topology by inflating all topological scales far beyond the current Hubble scale; but, absent evidence for a specific model of inflation, it is premature to discard the possibility of detecting a non-trivial topology of the Universe through an inflationary argument.

The case for considering cosmic topology is strengthened by the observation of several so-called large-angle anomalies in the cosmic microwave background (CMB) temperature data (see, e.g., \rcite{Planck:2013lks,Planck:2015igc,Planck:2019evm,Schwarz:2015cma,Abdalla:2022yfr} for reviews). These anomalies seem to indicate a preferred length scale in the Universe that is comparable to the current Hubble scale. 
If these anomalies are physical, i.e., not statistical flukes, then the anomaly scale has not been inflated beyond detectability, and could be the scale of cosmic topology.

A topologically non-trivial manifold can be represented by some \textit{fundamental domain} with a finite extent in at least one spatial dimension, and a prescription for how a geodesic intersecting a boundary surface of this domain resumes at a different boundary surface. 
These loops exhibit one or more characteristic length scales; if these length scales are sufficiently large compared to the Hubble scale, any observational consequences of the topology will fall below fundamental detection limits.

The most obvious observational consequence of such a topological structure would be the appearance of multiple images of distant cosmic objects: light coming from a far-away galaxy would arrive at our position via multiple different paths. Each manifold would exhibit a predictable  pattern of images of each astrophysical source to each observer
(that can depend on the position of each); hence the name cosmic crystallography \cite{Sokolov:1974,Fang:1983,Fagundes:1987,Lehoucq:1996qe,Roukema:1996cu,Weatherley:2003,Fujii:2011ga,Fujii:2013xsa} for the search for such patterns.
Detection of topology through such multiple images is complicated by the fact that different light paths lead to views of sources (such as galaxies) from different vantage points and with different look-back times, so both the morphology and time evolution of a source must be well understood to identify its multiple images. 
Since deep imaging over large sky areas became available, no  objects have been discovered with clones in well-separated locations on the sky \cite{Fujii:2013xsa}.
However, the observational difficulties and systematic uncertainties make precise quantitative constraints on topology from multiple-image limits difficult.

In 1996, Cornish, Spergel, and Starkman \cite{Cornish:1996kv,Cornish:1997ab} made a large conceptual step forward with their realization that topology would imprint a very specific signature in the CMB radiation. 
Most CMB photons travel to an observer unimpeded from near the last scattering surface, a spherical slice through the primordial photon distribution centered on the observer at a comoving distance of nearly the Hubble length. 
The insight was that, within a fundamental domain, the last scattering surface, comprised of all points that are at a fixed comoving distance from the observer along a geodesic of the spatial geometry, may no longer be a complete sphere, but that it will always be composed of spherical sections that intersect.

Any such intersection of spherical sections will be a circle if the metric is homogeneous and isotropic; at points along this circle, photons can travel from the last scattering surface in two different directions to arrive at one spacetime location. 
In a homogeneous universe, every place would have the same local density and temperature, and the topology would have no observable consequence.
But in a universe with small random spatial variations in physical conditions at the time of last scattering, a circle of spatial points encodes a specific pattern of fluctuations around a mean temperature. 
To the extent that CMB photons reflect the local conditions of the spacetime region from which they originated, an observer will see two separated circles in the microwave temperature sky that have the same pattern of hot and cold spots. 
The sizes, locations, and numbers of matched circle pairs depend on the specific manifold with its characteristic length scales, and for some topologies on the location of the observer within the manifold, but the existence of matched temperature circle pairs (for some observers) is a completely generic signature of any manifold having a length scale smaller than twice the comoving distance to the last scattering surface, so long as the geometry is sufficiently close to homogeneous and isotropic \cite{Cornish:1997ab,Cornish:1997hz,Cornish:1997rp}. 
But an observer will not necessarily see such circles --- only if there is a non-contractible closed loop through them that is shorter than the last scattering surface diameter $L_\mathrm{LSS}$.
Similar circles should then be found in polarization \cite{Riazuelo:2006tb} (although this case is subtler because polarization arises primarily from
primordial velocities, and the direction of the velocity vector relative to the photon direction depends on the photon path).

A generic search of the microwave sky for matched circles  is computationally challenging due to the large number of possible circle-pair centers, circle radii, and circle orientations. 
In addition, even perfect CMB sky maps, with no detector noise or foreground emission, have significant noise for this signal, arising from contributions to CMB anisotropies other than variations in the local gas temperature at last scattering.
These include the local gas velocity (``Doppler signal'') and the integrated Sachs-Wolfe (ISW) effect from time-varying gravitational potentials along the photon paths. 
Once full-sky microwave maps were created by the Wilkinson Microwave Anisotropy Probe (WMAP), full-sky matched-circle searches were completed, finding no matched circles in excess of those expected from noise \cite{deOliveira-Costa:2003utu,Cornish:2003db,ShapiroKey:2006hm,Mota:2010jb,Bielewicz:2010bh,Bielewicz:2011jz,Vaudrevange:2012da,Aurich:2013fwa}. 
This result was later confirmed using higher-resolution maps from the \textit{Planck} satellite \cite{ Planck:2013kqc,Planck:2015gmu,Starkman_Priv_Comm}. 
The null results imply that our Universe contains no closed space-like geodesics passing through our location with comoving proper length less than 98.5\% of $L_\mathrm{LSS}$.

This result has widely been interpreted to mean that any topologically non-trivial manifold with a characteristic length scale smaller than the observed limit is observationally ruled out. 
However, this is not the case. Even among the restricted class of spatially flat manifolds, there can be length scales significantly shorter than the diameter of the last scattering surface and still not violate the observational bounds. 
This is particularly true for topologies for which identification of boundary
surfaces involves a rotation or reflection.\footnote{
    In more mathematically precise language, in Euclidean manifolds where the covering space has been modded-out by a discrete subgroup of the $E^3$ isometry group that includes corkscrew motions or glide reflections --- i.e., all Euclidean manifolds except the simple 3-torus, the chimney space, the unrotated slab space, and the covering space --- the statement that the shortest closed spatial geodesic through us must be longer than $98.5\%$ of the diameter of the LSS does not imply that the shortest closed spatial geodesic through every point in the manifold must be longer than this.  
    This is possible because the topological boundary conditions break homogeneity: the length of the shortest closed geodesic through a particular point can vary substantially within the volume of the manifold.
}
In \rcite{COMPACT:2022gbl} we review the status of cosmic topology, arguing that past searches have been less thorough and their interpretation overly restrictive than typically thought. In particular, as we show here, the lower limit on the topology scale of an orientable Euclidean manifold can be a factor of 2 to 6 shorter than 
$L_\mathrm{LSS}$ (with even weaker limits possible in the rotated slab space).

In this paper, we systematically display constraints on the length scale of all spatially flat (Euclidean), orientable topologies from the condition that they do not violate the observed lack of circle pairs in the CMB temperature sky. 
In particular, we directly compute which observer locations in the fundamental domain result in a violation of the observed constraint, for a range of length scales in each topology. 
We do this by finding the comoving distance to the nearest observer clone in the covering space for observer locations throughout the manifold.
A particular topology with a given length scale  will have a fixed percentage of the fundamental domain volume allowed; the topology is disfavored if this percentage is too small, or strictly ruled out when it is zero. Our results highlight a number of cases where the minimum comoving length scale of allowed manifolds is significantly smaller than the comoving diameter of the last scattering surface.
The existence of many such topologies motivates further observational searches using different CMB signals, in particular the correlations between spherical harmonic components characteristic of a given topology \cite{COMPACT:2023rkp}. 

Observations of the CMB temperature and polarization imply that the mean spatial curvature is quite small, i.e., $\vert\Omega_K\vert\ll 1$, though perhaps $\Omega_K<0$ \cite{Planck:2018vyg}. 
Early-universe inflation, which can produce the basic properties of the observed Universe, is often taken to predict that spatial curvature should be near zero, but models that result in measurable curvature have also been considered. Furthermore, inflationary arguments would have to be reconsidered if cosmic topology were discovered, since the duration of inflation would then need to be short enough that the topology scale is not stretched to far beyond the current horizon scale. 
Despite these qualifications, for simplicity we take the background geometry of the Universe to be Euclidean (spatially flat) and orientable.
We will consider non-orientable Euclidean manifolds\footnote{
    Note that physical arguments about the
    consistency of quantum field theory may rule out non-orientable spaces
    \cite{Hawking:1973uf}, though we advocate caution in accepting the applicability of such no-go theorems, which, for example, assume that there are no topological features on small scales that would change the orientability of the manifold as encoded in the large-scale topology.
    } 
elsewhere, as we have identified new representations of their symmetries (loosely, new parameterizations of their possible fundamental domains) that merit detailed presentation. Non-flat geometries admit a countable infinity of topologies; in future work we will present limits on representative non-Euclidean manifolds from circle searches.

We make freely available a code that identifies the location of (and distance to) the nearest clone of any observer within a described fundamental domain. We also make available a second code that allows the user to establish whether specific parameters of an orientable Euclidean manifold are allowed or forbidden, depending on a chosen threshold of the fraction of observer locations producing matching circles. These codes\footnote{Available at \url{https://github.com/CompactCollaboration/EOT_FDSearch}.} will eventually be extended to include a wider set of manifolds.

\cref{sec:propManif} of this paper reviews relevant results about characterizing topology, fundamental domains, and covering spaces, and establishes our mathematical conventions. 
\cref{sec:methods} discusses how to compute distances to the nearest clones.
\cref{sec:topRest} displays results showing the portions of fundamental domains that are allowed for each topology considered as a function of the characteristic length scale of the fundamental domain given the observational constraint on matched circle pairs in the CMB temperature maps. 
In \cref{sec:conclusions}, we discuss how these results motivate further searches for topologies that are not ruled out by current observations, but which may still display detectable signatures more subtle than circle pairs. 
In the appendix we enumerate the specific set of topologies considered in this work.

\section{Properties of Orientable Euclidean Manifolds}\label{sec:propManif}

Each Euclidean manifold is characterized by a representation of its associated discrete, freely acting subgroup of the isometry group of \Espace --- the rotations, reflections, and translations in three dimensions.
In general, a representation is determined by a set of $d$ generators $g$ (where $d$ is the number of compact dimensions) 
that each acts on points $\vec{x}$ in the manifold as
\begin{equation}
    g:\vec{x}\to \mat{M} (\vec{x} - \vec{x}_0) + \vec{T} + \vec{x}_0\,. 
\end{equation}
Here $\mat{M}$ is an element of $O(3)$, i.e.,  a proper rotation (possibly the identity), a reflection, or some combination of both;
$\vec{T}$ is a translation vector; and $\vec{x}_0$ encodes the choice of origin\footnote{
    For a proper rotation, it is the position of an arbitrary point on the rotation axis with respect to the chosen origin; for a reflection, it is the position of an arbitrary point on the plane(s) of reflection; for the identity it is irrelevant.
}.
In this work we only consider orientable manifolds; thus $\mat{M}$ will be an element of $SO(3)$, i.e., either the identity or a proper rotation around an axis $\unitvec{n}$ by an angle $\alpha$, $\mat{R}_{\unitvec{n}}^\alpha$. 
The full complement of group elements is obtained by repeated application of the generators and their inverses. 

From the point of view of the covering space, each group element maps a given point to another distinct point in the manifold.
These points are referred to as ``clones.''
It is often convenient to think of the topologically non-trivial manifolds in terms of tilings of the covering space.
Each tile contains exactly one copy of each point --- i.e., no tile contains a point and any of its clones.
All tiles are exact copies of one another --- one can map one tile into any other by the application of some group element. 
The covering space is fully covered by the tiles --- i.e., no point in the covering space is omitted. 
One of these tiles is often referred to as the fundamental domain. 
However, it is worth recognizing that these tiles have no absolute physical meaning as one can group the points in space into tiles in many different ways --- consider, e.g., M.C. Escher's Flying Fish, a tiling of Euclidean 2-space by tiles with a very complex shape like a flying fish \cite{arthive}.
Thus the fundamental domain should not be misinterpreted as the shape of the Universe.
A convenient, and frequent, choice of fundamental domain is the Dirichlet domain of some particular point --- the set of all points in the manifold that are closer to that base point than to any of its clones.  
However, the shape of the Dirichlet domain typically depends on the choice of base point; thus it too should not be misinterpreted as the shape of the Universe.  
A given observer will have a Dirichlet domain shape which is not typically shared by other observers.
The physical characteristics of a topology are fully contained in the isometries of the 
manifold --- i.e., the generators and the group elements they generate.

The choice of topology fixes the matrices $\mat{M}$, leaving freedom to choose the translation vectors $\vec{T}$ in the generators.  
If $\mat{M}$ is a proper rotation, i.e., not the identity, then a redefinition of the origin $\vec{x}_0$ can be used to adjust the components of $\vec{T}$ that are perpendicular to the axis of rotation.
In each topology where this applies, we use this freedom to choose $\vec{T}$ parallel to the axis of rotation and relabel the coordinates so that $\vec{x}_0=\vec{0}$.  
Thus, for $\vec{x}=\vec{0}$, $g$ is always a pure translation, while for $\vec{x}\neq\vec{0}$ it is a so-called ``corkscrew'' motion.

There are six fully compact orientable Euclidean topologies: ``simple 3-torus,'' ``half-turn space,'' ``quarter-turn space,'' ``third-turn space,'' ``sixth-turn space,'' and ``Hantzsche-Wendt space'' (\E{1}--\E{6}). Two other orientable ``chimney spaces'' \E{11} and \E{12} are compact in two dimensions but not the third. Finally the ``slab space'' \E{16} is a manifold compact in just one dimension. 
The matrices $\mat{M}$ and the associated translations $\vec{T}$ for all of these orientable Euclidean topologies are given in the appendix. Further for all of these topologies (except for the special case of \E{16}, which we discuss in the appendix), there is, for each of the $\mat{M}$'s included in the set of group elements,
a smallest positive integer $N$ such that 
\begin{equation}
\label{eqn:genorder}
    \mat{M}^{N} = \identity\,, 
\end{equation}
where $\identity$ is the identity. This means that in \E{2}--\E{6} it is always possible to build three linearly independent pure translations out of products of the generators.
For example, $g^N$ (the application of the generator $g$ a total of $N$ times) is always a pure translation.
These three linearly independent translations serve as the generators of an ``associated \E{1}.'' Properly chosen, these translations generate the subgroup of the isometry group of the manifold that are all the pure translations.
Similarly, for \E{12}, we can construct an associated \E{11}.  
For each manifold, \E{1}--\E{6}, \E{11}, and \E{12}, the associated \E{1} or \E{11} is provided in the appendix.

\begin{figure}
  \centering
    \includegraphics[trim=6cm 0 0 0, width=0.45\textwidth]{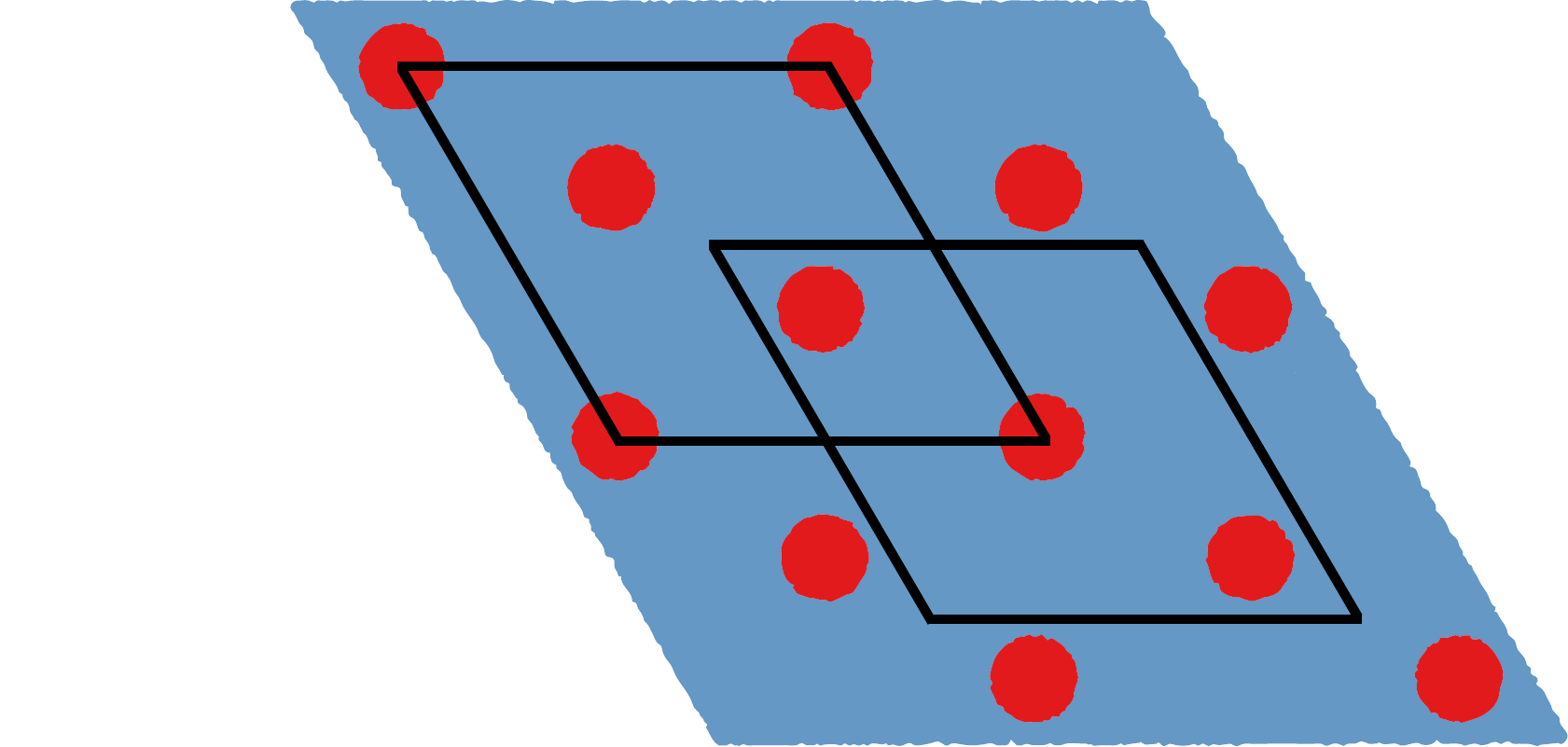}
  \caption{A representation of a section of the \E{4} covering space with parameter values $L_A = 2$ and $L_B = 0.95$ in units of $L_{\mathrm{LSS}}$, the comoving diameter of the last scattering surface. Observers who see matching circles are in red, and observers who do not see circles are in blue. The solid black lines represents two potential ``associated \E{1}'s'', depending on the choice of origin. The fraction of points in any associated \E{1} is equal, and in this case both \E{1} representations have a fraction of points allowed at 0.91.}
  \label{fig:figures_CoveringSpaceE4}
\end{figure}

These associated representations are of computational utility, as the natural fundamental domains constructed from those pure translations have simple shapes, allowing for the easy random selection of observers with uniform probability inside the associated \E{1} or \E{11}.
Uniform sampling of the associated \E{1} or \E{11} guarantees uniform sampling of the manifold.
A finite portion of an example covering space of \E{4} is shown in \cref{fig:figures_CoveringSpaceE4}, where every point represents an observer location and each black parallelogram encloses a choice of associated \E{1}. The red and blue points represent, respectively, whether or not an observer at that location would see matching circles in their sky. We do not provide a fundamental domain for each topology because it has no objective physical significance, nor do we provide a Dirichlet domain because it varies from observer to observer except in \E{1}, \E{11}, and an unrotated \E{16}.

\section{Clone Search}
\label{sec:methods}

The main goal of this work is to translate the non-detection of matched circle pairs in the CMB to constraints on the length scales of each compact, orientable, Euclidean topology.
For each topology and any set of topological parameter values characterizing a manifold with that topology, we determine the fraction of randomly distributed observers in that manifold who would have seen matched circle pairs.
If every observer in the manifold would have seen circles, then the circle searches based on the WMAP and \textit{Planck} data would completely rule out that manifold. In general, if a fraction $p$ of observers would see circles, then the manifold can be ruled out at approximately this $p$-value.

To achieve this goal, we randomly populate the manifold with observers, and determine the distance to each observer's nearest clone. For a given topology, we first choose a manifold based on the allowed range of parameters (see the appendix) and uniformly populate that manifold's associated \E{1} or \E{11} with potential observers.
Next, the eligible nearest clones of each observer are found by repeated application of generators. This list is composed of those clones lying in the region of the covering space adjoining the observer's associated \E{1} or \E{11}.
The distance to each of these clones is computed and the shortest distance, corresponding to the nearest clone, is noted.
If this shortest distance is less than $L_\mathrm{LSS}$ then the observer point is identified as an ``excluded point,'' otherwise it is an ``allowed point.''
This identification is made since an observer that has a clone closer than $L_\mathrm{LSS}$ will see matched circle pairs, violating the null result of the previous searches.\footnote{%
    More accurately, the searches did not find a value of the circle radii and center separations for which there was an excess of circle pairs exhibiting a value of the matching statistic that exceeded the  detection threshold for a 99.7\% C.L. detection, nor for a 95\% C.L. detection, for clone separations from the observer up to 98.5\% of $L_\mathrm{LSS}$. However, for slightly smaller distances, the chances of a false negative fall precipitously.
    } 
Once the associated \E{1} or \E{11} is sampled, the fraction $p$ of excluded points is determined and the manifold is classified based on the $p$-value: allowed ($p=0$), ruled out ($p=1$), or somewhere in between (all other cases are quantified by $p$).

The procedure for determining the closest clone to a point can be divided into two steps: applying non-trivial generators and applying pure translations.
The first step involves applying the non-trivial generators to the sampled point a number of times, up to the order of that generator in the associated \E{1}, described by $N$ in \cref{eqn:genorder}.
This step produces a set of clones, each with a unique configuration in the associated \E{1}.
For example, in \E{2}, we have two points in the set: the fiducial point and the point determined by a single application of $g^{\E{2}}_B$ (see \cref{eqn:E2generators} for this case and the appendix for the required generators of the other topologies). 
The second step involves applying the group of pure translations to each point in this new set of clones.
With the correct choice of translations, these two steps will give clones only within the $3\times3\times3$ cube of associated \E{1}, with the sampled associated \E{1} in the center.
The closest clone will always be in this small set of new points, and so drastically reduces the number of clones to be evaluated.
The minimum Euclidean distance of these clones is then found and compared to $L_\mathrm{LSS}$ to determine if the observer location is allowed or excluded.

\section{Topological Constraints from Observations}
\label{sec:topRest}

To examine how the distribution of allowed (or excluded) points in a manifold depends on their generators, we first fix the trivial generators --- those that are pure translations --- to a characteristic length scale greater than $L_{\mathrm{LSS}}$.
This is required since when the length scales of those trivial generators are smaller than the diameter of the last scattering surface every point in the manifold would have a clone close enough that matched circles would be observable.
By adjusting the remaining length scales, we are able to generate the regions of disallowed points for each topology and examine how those regions change as a function of the chosen length scales.

\begin{figure}[H]
  \centering
    \includegraphics[width = 0.9\textwidth]{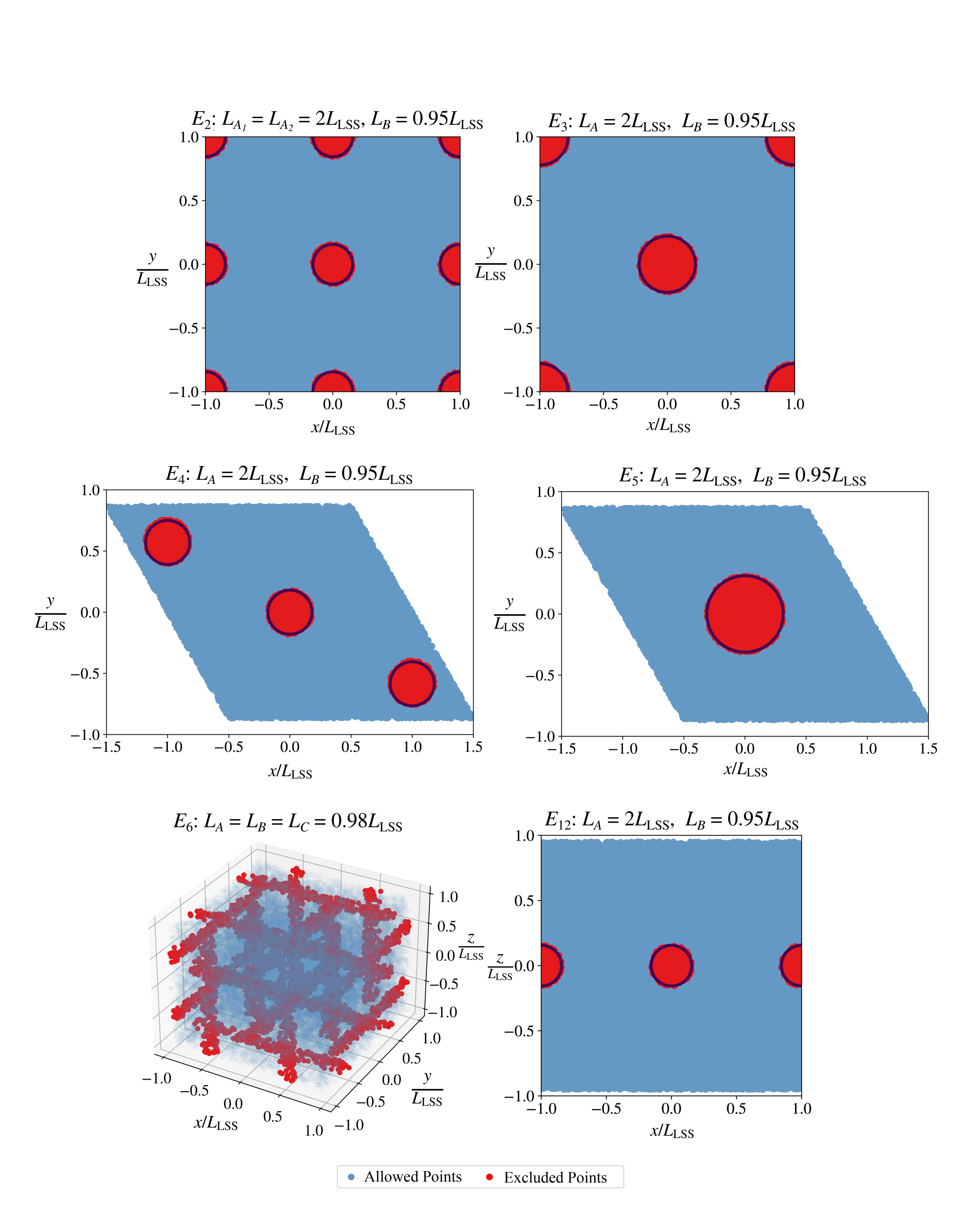}
  \caption{Classification of observer locations into excluded points represented in red (those who would see matched circle pairs), and allowed points represented in blue (those who would not)
  in six non-trivial, orientable topologies that have symmetry groups including corkscrew motions, i.e., \E{2}, \E{3}, \E{4}, \E{5}, \E{6}, and \E{12}. (\E{1} and \E{11} are homogeneous --- if any topology scale is less than $L_\mathrm{LSS}$, then all observers can see a matched circle pair, otherwise none can. \E{16} is discussed separately in the text.) The black circles indicate the analytical results for each of these topologies --- making use of the method described in \cref{eqn:analyticalManif} --- to compare to the numerical results. The coordinates $x$, $y$, $z$ denote the locations of the sampled observers relative to an origin on the axis of the corkscrew motion(s) of each manifold.
  The length scales associated with the fundamental domains are in units of $L_{\mathrm{LSS}}$. (See the appendix for details regarding the meaning of each length scale in each topology.) Parameters defining the  manifolds are chosen such that more than $90\%$ of the observers do not see matching circles in the sky. Note: \E{6} requires a choice of $L_{A} = L_{B} = L_{C} = 0.98L_{\mathrm{LSS}}$
    to reveal its symmetries.}
  \label{fig:figures_sampling}
\end{figure}

\cref{fig:figures_sampling} shows two-dimensional projections of manifolds with the topologies
\E{2}, \E{3}, \E{4}, \E{5}, and \E{12}
(along  the single corkscrew direction), and a three-dimensional plot of an \E{6} manifold, which has three corkscrew motions.
For the specific choices of the topology parameters of those manifolds indicated, 
observers in the blue locations would not see matched circle pairs, while observers in the red locations (which comprise  cylinders in the manifold, of which \cref{fig:figures_sampling} shows cross-sections) would see circle pairs.
For \E{2}, \E{3}, \E{4}, \E{5}, and \E{12},
the translations orthogonal to the corkscrew motion are chosen to be of length $L_\mathrm{LSS}$ --- large enough that the clones in those directions are too far to result in matched circle pairs.
The translations associated with the corkscrews are chosen such that just under 10\% of observers see circle pairs.
\E{1}  and \E{11} are not shown in the figure because they are homogeneous --- if any topology scale is less
than $L_\mathrm{LSS}$, then all observers can see a matched circle pair, otherwise none can. We discuss \E{16} below.

While the shape of the associated \E{1} and the pattern of excluded points are particular to each manifold, they do share common features. 
Most interestingly,  for appropriate parameters there are excluded cylindrical ``columns'' centered on each corkscrew axis (and its clones). This property is most clear in the three-dimensional plot of the circle-seeing (red) and non-circle-seeing (blue) regions of an \E{6} manifold in the bottom-left panel of \cref{fig:figures_sampling}. 
In \E{6}, each non-trivial generator ($g_A$, $g_B$, and $g_C$ as discussed in the appendix) has a corkscrew motion,
and we see the intersecting orthogonal cylinders in the figure.

For the simplest orientable topologies, the locations of these cylinders of excluded observers can be readily calculated analytically. For example, in the half-turn space, \E{2} (with corkscrew axis $\hat{z}$), 
for excluded points the shortest distance between clones satisfies
\begin{equation}
    \left\lVert\begin{pmatrix}x \\ y \\ z\end{pmatrix} - \begin{pmatrix}-x \\ -y \\ z + L_B\end{pmatrix}\right\rVert \leq L_{\mathrm{LSS}}\,,
\end{equation}
which reduces to the region
\begin{equation}
    \label{eqn:analyticalManif}
    x^2 + y^2 \leq \frac{L_{\mathrm{LSS}}^2-L_B^2}{4}\,.
\end{equation}
This is a cylinder with axis centered on the origin in the $xy$-plane, with radius $\frac{1}{2}\sqrt{L_{\mathrm{LSS}}^2-L_B^2}$.
Similar calculations can be performed for other topologies, which produce the boundaries shown by the black circles in \cref{fig:figures_sampling}. As expected, the radii of the excluded cylinders depend only on the length of the translation vector along the corkscrew axis.
   
\begin{figure}
  \centering
    \includegraphics[width=0.9\textwidth]{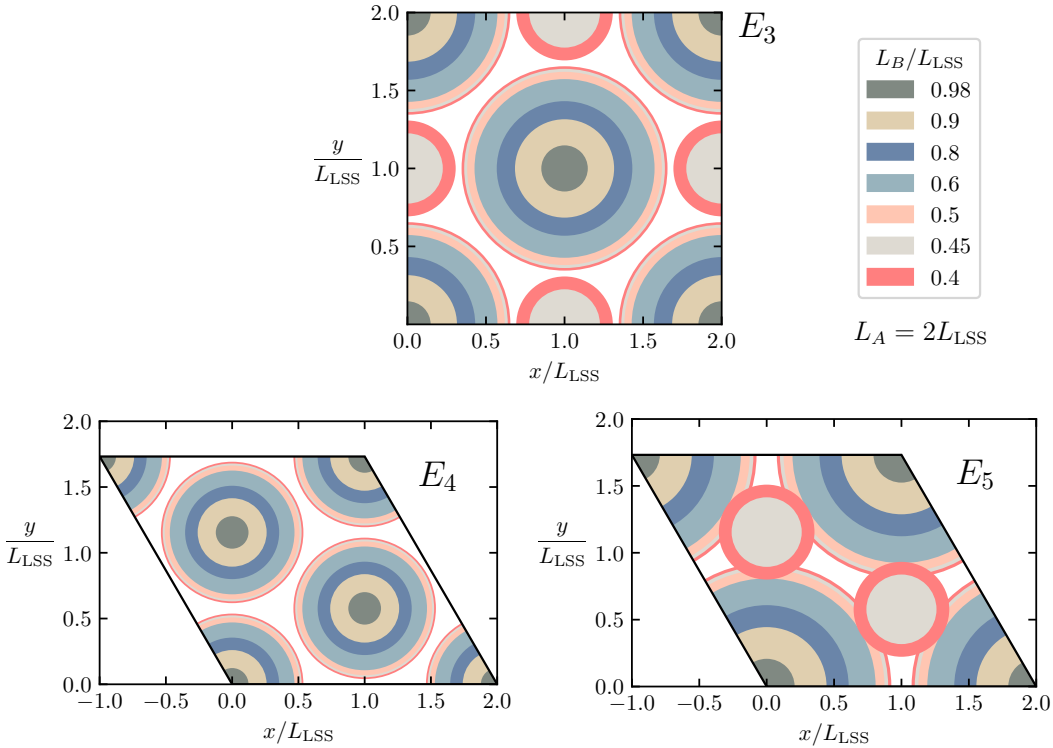}
  \caption{Locations $(x,y)$ of observers in planes of constant $L_B$ having clones closer than $L_{\mathrm{LSS}}$, the comoving diameter of the last scattering surface, as a function of $L_B$ for \E{3}, \E{4}, and \E{5} with $L_A = 2$ in units of $L_{\mathrm{LSS}}$. As the non-trivial length scale ($L_B$) decreases, the allowed region (in white) decreases. At $L_B \lesssim L_{\mathrm{LSS}}$, only a small fraction of observers see circle pairs.
  }
  \label{fig:figures_Contour}
\end{figure}

In \cref{fig:figures_Contour}, we make use of these analytical expressions and numerical results to show how the pattern of allowed (or excluded) points changes as the length scale $L_B$ (the length of the corkscrew motion about the $\hat{z}$ axis) is varied
in  \E{3}, \E{4}, and \E{5}.
$L_A$, the length of the two pure-translation generators in the $xy$-plane, is chosen to be $2L_\mathrm{LSS}$, so clones closer than $L_\mathrm{LSS}$ must be due, at least in part, to the corkscrew motion.
For all three topologies, increasing $L_B$ gradually reduces the fraction of observers with a clone closer than $L_{\mathrm{LSS}}$. We have restricted these contour plots to the topologies with two free parameters; however, similar figures can be produced for the other orientable topologies in a higher dimension parameter space.

We see in \cref{fig:figures_Contour} that as soon as $L_B$ is slightly less than $L_\mathrm{LSS}$ ($L_B=0.98 L_\mathrm{LSS}$ shown in dark grey), points in a disk centered at $(x,y)=(0,0)$ (and with any value of $z$) should see matched circle pairs and so are excluded. The radius of the disk depends on $L_B$, and grows with $\vert L_\mathrm{LSS} - L_B\vert$.
These are seen in all panels as partial disks centered on $(x,y)=(0,0)$, but also at the images of $(0,0)$ under the pure translations in the $xy$-plane --- i.e., at the corners of the parallelograms outlined in black.
However, another set of points is also excluded in \E{3} and \E{4} --- disks around points in the $xy$-plane whose images under the corkscrew motions have $x$ and $y$ coordinates that are clones of the original point.
For example, for \E{3}, the $\pi$-corkscrew takes $(L_\mathrm{LSS},L_\mathrm{LSS},z)$ to $(-L_\mathrm{LSS},L_\mathrm{LSS},z+L_B)$; but the application of one of the pure translations (by $(2L_\mathrm{LSS},0,0)$)  takes this to $(L_\mathrm{LSS},L_\mathrm{LSS},z+L_B)$, a distance $L_B$ directly above the original point. 
The same happens in \E{4} for observers at $(0,2L_\mathrm{LSS}/\sqrt{3},z)$ and $(L_\mathrm{LSS},L_\mathrm{LSS}/\sqrt{3},z)$; it does not happen in \E{5}.

In \E{3} and \E{5}, new excluded disks emerge when $L_B<L_\mathrm{LSS}/2$, centered at points where two applications of the corkscrew motion carry the disks to points that are clones of the original point.
For \E{3}, for example,
$(0,L_\mathrm{LSS},z)$ is taken to $(0,-L_\mathrm{LSS},z+2L_B)$, which can be taken back to $(0,L_\mathrm{LSS},z+2L_B)$ with a translation by $(0,2L_\mathrm{LSS},0)$. 
For \E{5} it is no surprise that the disks appear at the same locations where there are ``interior disks'' in \E{4}, since the corkscrew motion for \E{5} is a rotation by half the angle as in \E{4}.

\begin{figure}
  \centering
  \includegraphics[width=0.9\textwidth]{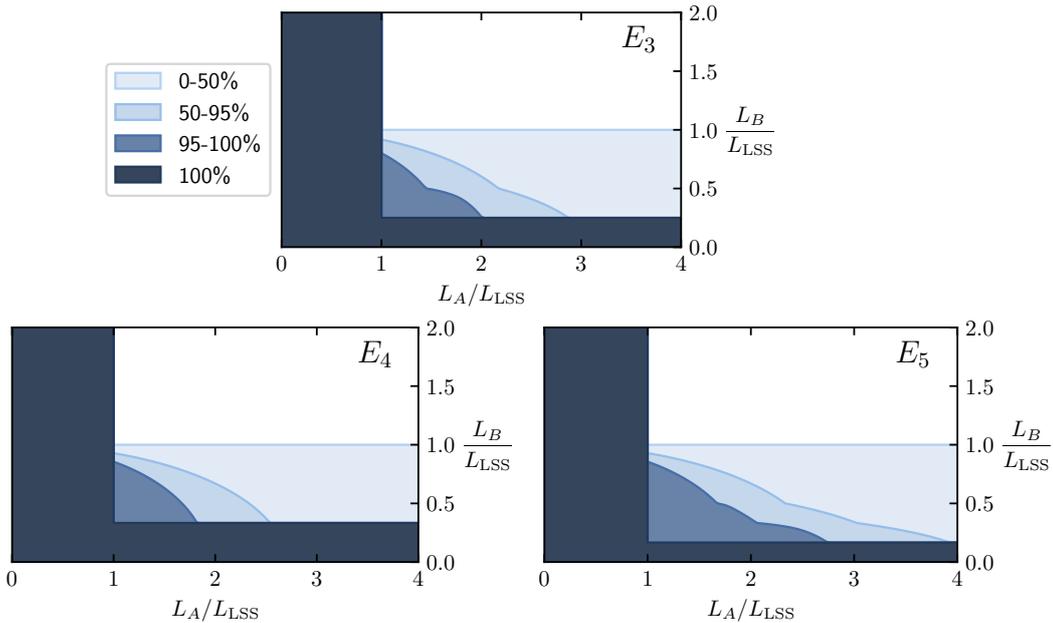}
  \caption{Excluded regions of the parameter space of the \E{3}, \E{4}, and \E{5} topologies, with the length scales measured in units of $L_\mathrm{LSS}$, the comoving diameter of the last scattering surface. Each region is limited by the percentage of observers having a close enough clone (closer than the $L_{\mathrm{LSS}}$) to produce a matched circle pair in the sky.
  For example, for any choice of parameters within the lightest gray region, 0 to $50\%$ of observers in the fundamental domain will have a clone close enough to produce a matched circle pair. 
  $L_A$ and $L_B$ are given in units of $L_\mathrm{LSS}$.
  }
\label{fig:figures_excludedmanifs}
\end{figure}

\begin{figure}
  \centering
    \includegraphics[width=0.65\textwidth]{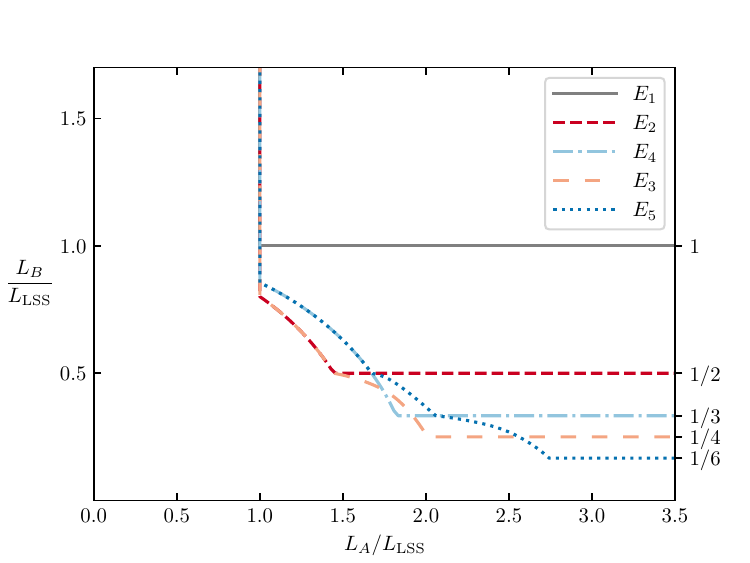}
\caption{
 Regions of topology parameter space where  observers would or would not see matched circle pairs. For each topology ($E_1$-$E_5$), for values of $L_A$ and $L_B$ in the regions below and to the left of the associated curve, $\geq95\%$ of observers have a clone closer than $L_\mathrm{LSS}$. (In the region to the right of $L_A/L_\mathrm{LSS}=1$ and above $L_B/L_\mathrm{LSS}=1$, no topology will produce any matched circles.)
    For the $E_2$ manifold we choose the right rectangular prism. 
    }
\label{fig:figures_exclusionCurveE1-5}
\end{figure}

To this point, we have explicitly chosen parameters such that only a small fraction ($<10\%$) of points in the manifold would have seen clones. We also aim to identify when a specific topology becomes either disfavored or strictly ruled out.
This can be done by examining the unique parameter space for each topology.
The dimensionality of this parameter space depends on the topology, and for the orientable Euclidean manifolds ranges from 2  to 6. In some topologies, we can straightforwardly calculate analytically the shortest distance to each observer's nearest clone,  and so determine how the constraint on the shortest distance around the Universe through us translates into a constraint on the parameters of that topology.  
In other topologies, the analytic approach is more fraught, or perhaps impossible.
Here, we choose to uniformly sample manifolds of each topology in order to establish the boundaries in parameter space between allowed regions (sufficiently low fraction of observers would see circles) and excluded regions (too high a fraction of observers would see circles).
That boundary depends on the fraction of circle-seeing observers one permits on the topology and on the topology parameters that specify the manifold.

The results of this sampling of the topology parameter spaces are shown in \cref{fig:figures_excludedmanifs}
for \E{3}, \E{4}, and \E{5} --- the three manifolds with just two topology parameters.
In the most strict case --- no observer sees circles --- we have sharp cutoffs defined by the condition $L_{\mathrm{max}} = L_\mathrm{LSS}$,
where $L_{\mathrm{max}}$ is the length of the longest  translation vector of the three generators.\footnote{
    There is an important subtlety here: one must have chosen the three generators with the shortest possible translation vectors. The parameterizations presented in the appendix make that choice.
    }
This makes intuitive sense --- for pure translations, the distance to the nearest clone is the length of the translation. 
For corkscrew motions, 
the points along the axis of rotation experience a pure translation; they
are the closest to their clones, 
and only become excluded when the pure translation does not produce matching circles.
If we allow a growing percentage of observers to see circles, we begin to expand the parameter space, until we reach the other  limit: manifolds in which no observer sees circles. These results are summarized in \cref{fig:figures_exclusionCurveE1-5} for the case where at least 95\% of observers in the manifold see matched circles. We include \E{2} (which has four parameters) with $L_{\A{1}}=L_{\A{2}} \equiv L_A$ and $\alpha=\pi/2$ (see 
the appendix for details of the \E{2} parameters).

\begin{figure}[H]
  \centering
    \includegraphics[width=1\textwidth, keepaspectratio]{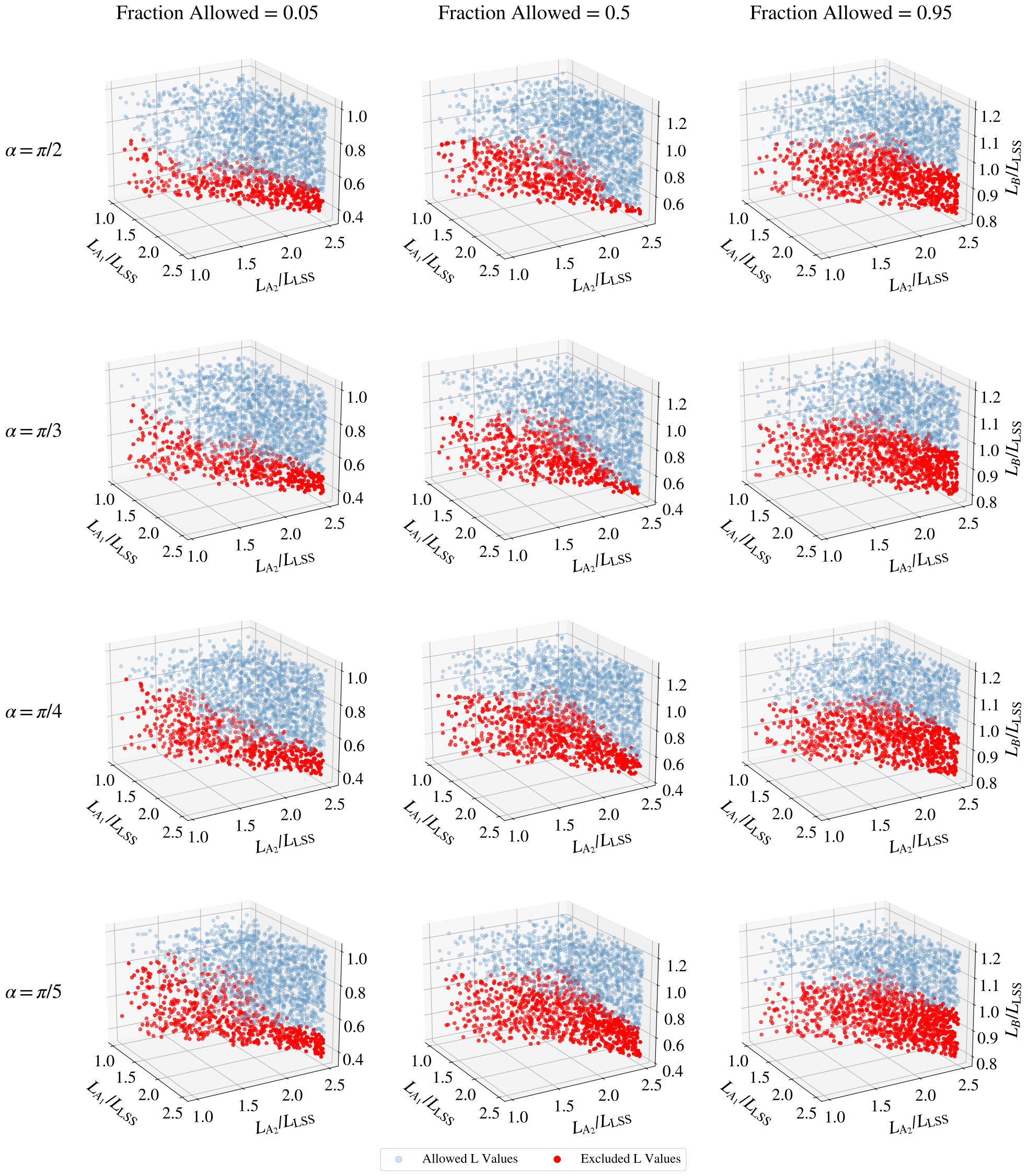}
  \caption{
  Constraints on parameter space of the \E{2} topology with length scales $L_{\A{1}}$, $L_{\A{2}}$, and $L_B$ measured in units of $L_\mathrm{LSS}$, the comoving diameter of the last scattering surface. Every point within a graph describes a unique representation of an \E{2} fundamental domain. Left to right: the `fraction threshold' describing how many points within the fundamental domains must not see circles in order for the parameter values to be considered excluded. Top to bottom: the angle $\alpha$ of the fundamental domains.}
  \label{fig:figures_E2Param}
\end{figure}

\begin{figure}
  \centering
    \includegraphics[width = 0.95\textwidth,keepaspectratio]{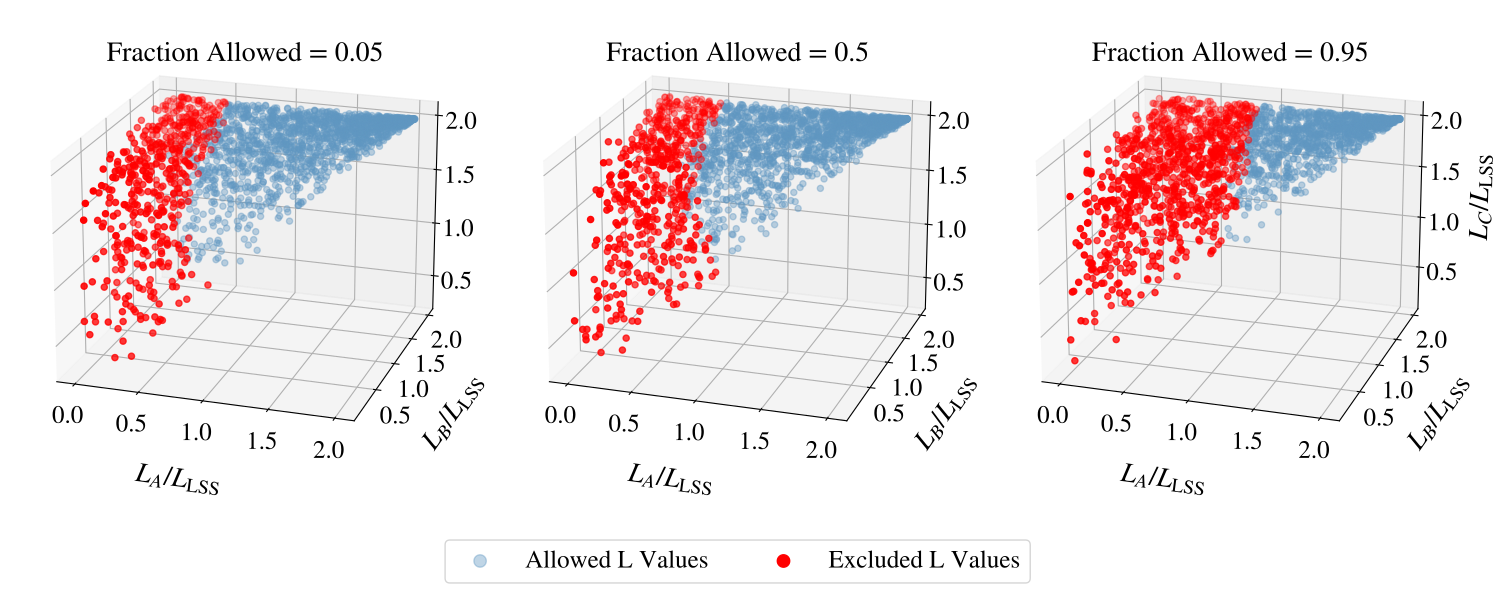}
  \caption{As in \cref{fig:figures_E2Param}, but for the \E{6} topology with length scales $L_{A}$, $L_{B}$, and $L_{C}$. 
  }
  \label{fig:figures_E6Param}
\end{figure}

Topologies with more than two free parameters, i.e., \E{2} and \E{6}, are more difficult to produce analytic results for, and so we rely on numerical visualizations of the three-dimensional parameter space. 
\cref{fig:figures_E2Param} and \cref{fig:figures_E6Param} show constraints on the parameter spaces for these two topologies. 
More precise results for values near the boundary can be identified using the provided associated \E{1} sampling code on GitHub.

\E{16} is a special case. If the rotation angle $\zeta$ associated with the corkscrew is zero (see section \ref{secn:topologyE16} of the appendix), then it is like \E{1} and \E{11}. 
If $\zeta\neq0$ and $L_\mathrm{LSS}<L$, then no observer sees a circle.
If $\zeta\neq0$ and $L_\mathrm{LSS}/2<L<L_\mathrm{LSS}$, then only observers in a cylinder around the rotation axis would see circles;
the larger $\vert\zeta\vert$ is, the larger the radius of the cylinder.
If $L_\mathrm{LSS}/3<L<L_\mathrm{LSS}/2$, then the cylinder of observers who see circles grows --- the closer $\zeta$ is to $\pm\pi$, the larger the radius;
if  $\zeta= \pm \pi$, 
then all observers would see circles.
A similar transition happens for $L_\mathrm{LSS}/4<L<L_\mathrm{LSS}/3$ --- a larger cylinder of circle-observing observers, with the radius determined by how close $\zeta$ is to $2p\pi/3$, for $p=\pm1$ or $\pm2$. 
And so on, as $L$ falls below $L_\mathrm{LSS}/q$ for each successive positive integer $q$.

\section{Conclusions}
\label{sec:conclusions}

We have shown herein the limits on the parameters of the orientable Euclidean 
topologies from non-observation of matching circle pairs in the CMB sky are much less restrictive than previously assumed. 
For the \E{1} and \E{11} topologies,  
the shortest closed spatial geodesic around the Universe $L_\mathrm{min}$,
which is also the length of the shortest of the  translations associated with the topology, must be longer than approximately $L_\mathrm{LSS}$.
However, in \E{2}--\E{6}, \E{12}, and \E{16} 
the constraint on $L_\mathrm{min}$ is weaker by a topology-dependent factor that ranges from 2 in \E{2} and \E{12} to 6 in \E{5}, and an arbitrarily large factor in \E{16}. Sampling of observer locations in each manifold has been used to identify the allowed and excluded regions within each manifold and to place constraints on the parameter space of each topology.

We have described a method for identifying the location of the nearest clone in the covering space of $E^3$, depending on the chosen parameters of a manifold with a given topology. We demonstrated in \cref{fig:figures_Contour} how the regions of a manifold in which observers see matched circle pairs in the CMB sky depend on the parameters of the topology. 
We have also provided instructive visualisations of the allowed and excluded parameters of each topology. Tools will be made publicly available on GitHub for finding the location and distance to the nearest clone; calculating the fraction of observer locations in a given manifold with matched circle pairs; and searching parameter space of a topology at a specified exclusion threshold.

We have considered only orientable Euclidean topologies, ignoring the non-flat and non-orientable cases. In future studies we will constrain the parameter spaces for those topologies, and provide additional descriptions of their properties.
An upcoming paper \cite{COMPACT:2023rkp} will provide more details about the properties of the Euclidean manifolds, both orientable and non-orientable, including detailed descriptions of the most general parameterizations of the manifolds of each topology, 
and present the resulting eigenmodes of the scalar Laplacian.
These are basic building blocks necessary to further the generic search for non-trivial cosmic topology.

Certainly the available parameter space in which to search for cosmic topology is considerably more open than previously believed. In \rcite{COMPACT:2023rkp}, we will show that the accessible microwave background information available due to any non-trivial topology 
can also be larger than previously understood.

\acknowledgments
We thank Jeffrey Weeks and David Singer for valuable conversations.
Y.A. acknowledges support by the Richard S. Morrison Fellowship, from research projects PGC2018-094773-B-C32 and PID2021-123012NB-C43, by the Spanish Research Agency (Agencia Estatal de Investigaci\'on)'s grant RYC2020-030193-I/AEI/10.13039/501100011033 and the European Social Fund (Fondo Social Europeo) through the  Ram\'{o}n y Cajal program within the State Plan for Scientific and Technical Research and Innovation (Plan Estatal de Investigaci\'on Cient\'ifica y T\'ecnica y de Innovaci\'on) 2017-2020, and by the Spanish Research Agency through the grant IFT Centro de Excelencia Severo Ochoa No CEX2020-001007-S funded by MCIN/AEI/10.13039/501100011033. C.J.C., A.K. and G.D.S.  acknowledge partial support from NASA ATP grant RES240737; G.D.S. from DOE grant DESC0009946; P.P., Y.A., G.D.S., O.G. and S.S. from the Simons Foundation; Y.A., A.H.J. and G.D.S. from the Royal Society (UK); and A.H.J. from STFC in the UK.  A.T. is supported by the Richard S. Morrison Fellowship. J.R.E. acknowledges support from the European Research Council under the Horizon 2020 Research and Innovation Programme (Grant agreement No.~819478).

\appendix

\section{Generators of Orientable Euclidean Topologies}

\label{app:generators}

This appendix collects expressions for the generators of the topologies considered in this paper, most of which have appeared previously in the literature though we do provide explicit expressions for the most general form of the \E{2} generators which had previously only been suggested (see, e.g., \rcite{Riazuelo:2003ud}).
Also the possibility of a corkscrew motion rather than a pure translation appears to have been overlooked in \E{16}. In \rcite{COMPACT:2023rkp}, we will find that for non-orientable Euclidean topologies, the possible manifolds are considerably more diverse than previously expected.
 
These generators depend on one or more free parameters.
Our parameter sets are chosen to represent all possible cases while simplifying the form of the generators. 
We provide restrictions on the parameter values to avoid duplicate representations of the same physical space, defined by identical patterns of clones. The rationale behind these conditions on the parameters is given in \rcite{COMPACT:2023rkp}.

\subsection{\E{1}: Simple 3-torus}

The simple 3-torus is the most basic compact Euclidean manifold, only containing translations in its symmetry group. We use the freedom to orient the observer to align the shortest translation along the $x$-direction, and then to set the $z$-component of the second shortest translation to zero.
Equivalently, this means that two translation vectors $\vec{T}^{\E{1}}_{\A{1}}$ and $\vec{T}^{\E{1}}_{\A{2}}$ lie in the $xy$-plane and the third $\vec{T}^{\E{1}}_{\A{3}}$ has a non-zero $z$-component.
The generators of the simple 3-torus are then given in the usual orthogonal Cartesian basis by
\begin{align} 
    \label{eqn:E1generators}
    \mat{M}_A^{\E{1}} &= \identity
     \nonumber\,, \\
    \vec{T}^{\E{1}}_{\A{1}} &= L_{\A{1}}(1,\,0,\,0) \equiv \vec{T}_1^{\E{1}}\,, \\
    \vec{T}^{\E{1}}_{\A{2}} &= L_{\A{2}}(\cos\alpha,\,\sin\alpha,\,0) \equiv \vec{T}_2^{\E{1}}\,, \nonumber\\
    \vec{T}^{\E{1}}_{\A{3}} &= L_{\A{3}}(\cos\beta\cos\gamma,\,\cos\beta\sin\gamma,\,\sin\beta) \equiv \vec{T}_3^{\E{1}}\,. \nonumber
\end{align}
Here, $\alpha$ is the angle between the two vectors in the $xy$\nobreakdash-plane, and $\beta$ and $\gamma$ are the angles describing the orientation of the third vector. We take $L_{\A{1}}$, $L_{\A{2}}$, and $L_{\A{3}}$ as the ``length scales'' of each of these translation vectors, which in this simple case correspond to side lengths of the fundamental domain.
This choice simply defines a convenient coordinate system for any set of three linearly independent vectors in three dimensions.

Note that we have introduced indices on the matrices $\mat{M}$, the translations $\vec{T}$, and the lengths associated with each generator.
We label each distinct matrix within a given topology as $\mat{M}^{\E{i}}_a$, $a \in \{A,B,C\}$.
For \E{1} all the matrices are the same, $\mat{M}^{\E{1}}_{A}$.
For each fixed $a$, we label the translations as $\vec{T}^{\E{i}}_{a_j}$ and the lengths as $L_{a_j}$, where the additional index $j\in\{1, 2, 3\}$ is only included when required.
For \E{1}, the three vectors associated with $\mat{M}^{\E{1}}_{A}$ are thus  $\vec{T}^{\E{i}}_{\A{1}}$, $\vec{T}^{\E{i}}_{\A{2}}$, and $\vec{T}^{\E{i}}_{\A{3}}$. 

\subsection{\E{2}: Half-turn space}
\label{secn:topologyE2}

The half-turn space replaces one of the translations in \E{1} with a corkscrew motion by $\pi$.
We have used the freedom to rotate the coordinate system to set the axis of the associated rotation  $\mat{M}_B^{\E{2}}$ parallel to $\unitvec{z}$, and then to set the $y$-component of the shorter of the two pure translations to zero, i.e., $\vec{T}^{\E{2}}_{\A{1}y}=0$.
(The  orthogonality of the pure translations to the axis of rotation is not a choice, but a necessity \cite{COMPACT:2023rkp}.)
The generators of the half-turn space are then given by
\begin{align}
    \label{eqn:E2generators}
    \mat{M}^{\E{2}}_A &= \identity\,,  
       \nonumber \\
    \vec{T}^{\E{2}}_{\A{1}} &= L_{\A{1}}(1,\,0,\,0) \equiv \vec{T}^{\E{2}}_1\,, \nonumber \\
    \vec{T}^{\E{2}}_{\A{2}} &= L_{\A{2}}(\cos\alpha,\,\sin\alpha,\,0)
    \equiv \vec{T}^{\E{2}}_2\,; \\
    \mat{M}^{\E{2}}_B &= \mat{R}_{\unitvec{z}}^{-\pi} = \diag(-1,-1,1)\,,  \nonumber \\
    \vec{T}^{\E{2}}_{B} &= L_B (0,\,0,\,1)\,. \nonumber
\end{align}
The associated \E{1} is given by the translation vectors $\vec{T}^{\E{2}}_1$ and $\vec{T}^{\E{2}}_2$, and a third translation vector $\vec{T}^{\E{2}}_3$ obtained from $(g_B^{\E{2}})^2$. Explicitly,
\begin{equation}
    (g_B^{\E{2}})^2: \vec{x} \to  \vec{x} + (\identity + \mat{M}^{\E{2}}_B) \vec{T}^{\E{2}}_{B}\,,
\end{equation}
from which we define
\begin{equation}
    \label{eqn:TE2_3}
    \vec{T}^{\E{2}}_3 \equiv (\identity + \mat{M}^{\E{2}}_B) \vec{T}^{\E{2}}_{B} = 2L_B (0,\,0,\,1)\,. 
\end{equation}
The following restrictions on the \E{2} parameters ensure we do not have
duplicate representations of the same physical space:
\begin{enumerate}
    \item $0<L_{\A{1}}$\,; 
    \item $0< L_{\A{1}}/L_{\A{2}} \leq 1$\,, \quad $0 < L_{\A{1}}/L_B < \infty $\,; 
    \item $\vert\cos\alpha\vert < L_{\A{1}}/ (2 L_{\A{2}})$\,. 
\end{enumerate}

\subsection{\E{3}: Quarter-turn space}
\label{secn:topologyE3}
The quarter-turn space replaces the $\pi$ corkscrew of \E{2} with a corkscrew motion by $\pi/2$.
The same freedom to rotate the coordinate system as in \E{2} has been used to set the axis of rotation of $\mat{M}_B^{\E{3}}$ parallel to $\unitvec{z}$, and then to set the $y$-component of one of the two pure translations to zero, i.e., $\vec{T}^{\E{3}}_{\A{1}y}=0$. 
(The  orthogonality of the pure translations to the axis of rotation is not a choice, but a necessity \cite{COMPACT:2023rkp}, as is the equal length of the two pure translations and their orthogonality.)
The generators of the quarter-turn space are then given by
\begin{align}
    \mat{M}^{\E{3}}_A &= \identity\,,  \nonumber\\
    \vec{T}^{\E{3}}_{\A{1}} &= L_A (1,\,0,\,0) \equiv \vec{T}^{\E{3}}_1\,,  \nonumber\\
    \vec{T}^{\E{3}}_{\A{2}} &= L_A (0,\,1,\,0) \equiv \vec{T}^{\E{3}}_2\,; \\
    \mat{M}_B^{\E{3}} &= R_{\unitvec{z}}^{-\pi/2} =
        \begin{pmatrix}
        0 & 1 & 0\\
        -1 & 0 & 0\\
        0 & 0 & 1
        \end{pmatrix}\,, 
        \nonumber
        \\
    \vec{T}^{\E{3}}_{B} &= L_B (0,\,0,\,1)\,. 
    \nonumber
\end{align}
The associated \E{1} is given by the translation vectors $\vec{T}^{\E{3}}_1$, $\vec{T}^{\E{3}}_2$, and a third translation vector $\vec{T}^{\E{3}}_3$ determined from $(g_B^{\E{3}})^4$ in a fashion analogous to \cref{eqn:TE2_3},

\begin{equation}
    \vec{T}^{\E{3}}_3 \equiv 4L_B (0,\,0,\,1)\,.
\end{equation}
The following restrictions on the \E{3} parameters ensure we do not have
duplicate representations of the same physical space:
\begin{enumerate}
    \item $0<L_A$\,; 
    \item $0< L_A / L_B < \infty $\,. 
\end{enumerate}

\subsection{\E{4}: Third-turn space}
\label{secn:topologyE4}
The third-turn space replaces the $\pi$ corkscrew of \E{2} with a corkscrew motion by $2\pi/3$.
The same freedom to rotate the coordinate system as in \E{2} can be used to set the axis of rotation of $\mat{M}_B^{\E{4}}$ parallel to $\unitvec{z}$, 
and then to set the $y$-component of one of the two pure translations to zero, i.e., $\vec{T}^{\E{4}}_{\A{1}y}=0$. 
(The  orthogonality of the pure translations to the axis of rotation is not a choice, but a necessity \cite{COMPACT:2023rkp}, as is the equal length of the two pure translations and the angle between them.)
The generators of the third-turn space are then given by
\begin{align}
    \mat{M}^{\E{4}}_A &= \identity\,, \nonumber\\
    \vec{T}^{\E{4}}_{\A{1}} &= L_A (1,\,0,\,0) \equiv \vec{T}^{\E{4}}_1\,,\\
    \vec{T}^{\E{4}}_{\A{2}} &= R_{\unitvec{z}}^{-2\pi/3} \vec{T}^{\E{4}}_{\A{1}}  = L_A (-1/2,\,\sqrt{3}/2,\,0) \equiv\vec{T}^{\E{4}}_2\,;\nonumber\\
    \mat{M}_B^{\E{4}} &= \mat{R}_{\unitvec{z}}^{-2\pi/3} =
        \begin{pmatrix}
        -1/2 & \sqrt{3}/2 & 0\\
        -\sqrt{3}/2 & -1/2 & 0\\
        0 & 0 & 1
        \end{pmatrix}\,,
        \nonumber
        \\
    \vec{T}^{\E{4}}_{B} &= L_B (0,\,0,\,1)\,.\nonumber
\end{align}
The associated \E{1} is given by the translation vectors $\vec{T}^{\E{4}}_1$ and $\vec{T}^{\E{4}}_2$, and a third translation vector $\vec{T}^{\E{4}}_3$ determined from $(g_B^{\E{4}})^3$, as in \E{2} and \E{3},
\begin{equation}
    \vec{T}^{\E{4}}_3 \equiv 3L_B(0,\,0,\,1)\,.  
\end{equation}
The following restrictions on the \E{4} parameters ensure we do not have
duplicate representations of the same physical space:
\begin{enumerate}
    \item $0<L_A$\,;
    \item $0< L_A / L_B < \infty $\,.
\end{enumerate}

\subsection{\E{5}: Sixth-turn space}
\label{secn:topologyE5}

The sixth-turn space replaces the $\pi$ corkscrew of \E{2} with a corkscrew motion by $\pi/3$.
The same freedom to rotate the coordinate system as in \E{2} has been used to set the axis of rotation of $\mat{M}_B^{\E{5}}$ parallel to $\unitvec{z}$, and then to set the $y$-component of one of the two pure translations to zero, i.e., $\vec{T}^{\E{5}}_{\A{1}y}=0$. 
(The  orthogonality of the pure translations to the axis of rotation is a necessity \cite{COMPACT:2023rkp}, as is the equal length of the two pure translations and the angle between them.)
The generators of the sixth-turn space are then given by
\begin{align}
    \mat{M}^{\E{5}}_A &= \identity\,, \nonumber\\
    \vec{T}^{\E{5}}_{\A{1}} &= L_A (1,\,0,\,0)\,,\nonumber\\
    \vec{T}^{\E{5}}_{\A{2}} &= L_A (-1/2,\,\sqrt{3}/2,\,0)\,;\\
    \mat{M}_B^{\E{5}} &= \mat{R}_{\unitvec{z}}^{-\pi/3} =
        \begin{pmatrix}
        1/2 & \sqrt{3}/2 & 0\\
        -\sqrt{3}/2 & 1/2 & 0\\
        0 & 0 & 1
        \end{pmatrix}\,,
        \nonumber
        \\
    \vec{T}^{\E{5}}_{B} &= L_B (0,\,0,\,1)\,.\nonumber
\end{align}
The associated \E{1} is given by the translation vectors $\vec{T}^{\E{5}}_1$ and $\vec{T}^{\E{5}}_2$, and a third translation vector $\vec{T}^{\E{5}}_3$ determined from $(g_B^{\E{5}})^6$,
\begin{equation}
    \vec{T}^{\E{5}}_3 \equiv 6L_B(0,\,0,\,1)\,. 
\end{equation}
The following restrictions on the \E{5} parameters ensure we do not have
duplicate representations of the same physical space:
\begin{enumerate}
    \item $0<L_A$\,;
    \item $0< L_A / L_B < \infty $\,.
\end{enumerate}

\subsection{\E{6}: Hantzsche-Wendt space}

\label{secn:topologyE6}
The Hantzsche-Wendt space replaces each of the translations of \E{1} with a corkscrew motion by angle $\pi$.
We have used the freedom to rotate the coordinate system to set the axes of rotation of these three corkscrews to be parallel to the three coordinate axes.
Their orthogonality to one another has been proven \cite{Thurston1982ThreeDM}.
The appropriate parameterization of the associated translations is demonstrated in \rcite{COMPACT:2023rkp}.
The generators of the Hantzsche-Wendt space are then given by
\begin{align}
    \mat{M}^{\E{6}}_{A} &= \mat{R}_{\unitvec{x}}^{-\pi} = \diag(1,-1,-1)\,, \nonumber\\
    \vec{T}^{\E{6}}_{A} &= (L_{A},\, L_{B},\, 0)\,;\nonumber\\
        \mat{M}^{\E{6}}_{B} &= \mat{R}_{\unitvec{y}}^{-\pi} = \diag(-1,1,-1)\,, \\
    \vec{T}^{\E{6}}_{B} &= (0,\, L_{B},\, L_{C})\,;\nonumber\\
        \mat{M}^{\E{6}}_{C} &= \mat{R}_{\unitvec{z}}^{-\pi} = \diag(-1,-1,1)\,, \nonumber\\
    \vec{T}^{\E{6}}_{C} &= (L_{A},\, 0,\, L_{C})\,. \nonumber
\end{align}
The associated \E{1} is given by the translation vectors generated from $(g_a^{\E{6}})^2$ for $a\in\{A, B, C\}$,
\begin{align}
   \vec{T}^{\E{6}}_1 &\equiv (2L_{A},\,0,\,0)\,, \nonumber \\
   \vec{T}^{\E{6}}_2 &\equiv (0,\,2L_{B},\,0)\,, \\
   \vec{T}^{\E{6}}_3 &\equiv (0,\,0,\,2L_{C})\,. \nonumber
\end{align}
The following restrictions on the \E{5} parameters ensure we do not have
duplicate representations of the same physical space:
\begin{enumerate}
    \item $0<\vert L_{A}\vert$\,,
    \item $1 \geq \vert L_{A}/L_{B} \vert \geq  \vert L_{A}/L_{C}\vert$\,.
\end{enumerate}

\subsection{\E{11}: Chimney space}
\label{secn:topologyE11}
The chimney space has only two compact dimensions, and has only translations in its group.
We take the shorter of the two translations to be in the $x$-direction, and then set the $z$-component of the other translation to zero.
The generators of the chimney space are then given by
\begin{align} 
    \label{eqn:E11generators}
    \mat{M}_A &= \identity\,,\nonumber\\
    \vec{T}^{\E{11}}_{\A{1}} &= L_{\A{1}}(1,\,0,\,0) \equiv \vec{T}^{\E{11}}_1\,,\\
    \vec{T}^{\E{11}}_{\A{2}} &= L_{\A{2}}(\cos\alpha,\,\sin\alpha,\,0) \equiv \vec{T}^{\E{11}}_2\,.\nonumber
\end{align}
Much like \E{1}, these two translations can be taken to define the fundamental domain.

\subsection{\E{12}: Chimney space with half turn}
\label{secn:topologyE12}

The chimney space with a half turn replaces one of the two translations of \E{11} with a corkscrew motion by $\pi$.
We choose the axis of rotation of $\mat{M}_B^{\E{12}}$ parallel to $\unitvec{y}$, and then set the $z$-component of the pure translation to zero, i.e., $\vec{T}^{\E{2}}_{\A{1}z}=0$. 
The generators of the chimney space with a half turn are then given by
\begin{align}
    \mat{M}^{\E{12}}_{A} &= \identity\,, \nonumber\\
    \vec{T}^{\E{12}}_{A} &= L_A(\cos\alpha,\,\sin\alpha,\,0)\equiv \vec{T}^{\E{12}}_1\,;\\
    \mat{M}^{\E{12}}_{B} &= \mat{R}_{\unitvec{y}}^{-\pi} = \diag(-1, 1, 1)\,, \nonumber \\
    \vec{T}^{\E{12}}_{B} &= L_B(\cos\beta\cos\gamma,\,\cos\beta\sin\gamma,\,\sin\beta)\,. \nonumber
\end{align}
The associated \E{11} is given by the translation vectors $\vec{T}^{\E{12}}_1$ and the second determined from $(g_B^{\E{12}})^2$,
\begin{equation}
    \vec{T}^{\E{12}}_2 \equiv (0,\,2L_B\cos\beta\sin\gamma,\,0)\,.
\end{equation}

\subsection{\E{16}: Slab space with rotation}
\label{secn:topologyE16}

The slab space is compact in only one dimension.  
All group elements are integer repetitions of the generator (or its inverse). 
We can always choose the compact dimension to be in the $z$-direction.
The standard representation of \E{16} is a pure translation.
However, the pure translation is a special case as it is homogeneous --- all observers see the same pattern of clones.
An additional possibility is to also include a corkscrew motion by an arbitrary angle $\zeta$.
This possibility in the slab space appears to be new, at least in the cosmology literature.
The generator that includes a corkscrew motion is given by
\begin{align} 
    \mat{M}^{\E{16}}(\zeta) &= \mat{R}_{\unitvec{z}}^{\zeta}\,,\qquad 0\leq\zeta < 2\pi\,, \nonumber \\
    \vec{T}^{\E{16}} &=  L(0,\, 0,\, 1)\,, \qquad 0< L\,.
\end{align}
For $\zeta\neq0$ we have  changed the origin to eliminate the $x$ and $y$ components of $\vec{T}^{\E{16}}$; for $\zeta=0$ we have set $\vec{T}^{\E{16}}\|\hat{z}$ by rotating the coordinate system.

This is a strange case, see \cref{sec:topRest} (and \rcite{COMPACT:2023rkp} for full details).

\bibliographystyle{utphys}
\bibliography{topology,additional}

\providecommand{\href}[2]{#2}\begingroup\raggedright\begin{thebibliography}{10}

\bibitem{Lachieze-Rey:1995}
M.~{Lachieze-Rey} and J.~{Luminet}, ``{Cosmic topology},''
  \href{http://dx.doi.org/10.1016/0370-1573(94)00085-H}{{\em Phys. Rep.}
  {\bfseries 254} (Mar., 1995) 135--214},
  \href{http://arxiv.org/abs/gr-qc/9605010}{{\ttfamily arXiv:gr-qc/9605010
  [gr-qc]}}.

\bibitem{HAWKING1978349}
S.~Hawking, ``Spacetime foam,''
  \href{http://dx.doi.org/https://doi.org/10.1016/0550-3213(78)90375-9}{{\em
  Nuclear Physics B} {\bfseries 144} no.~2, (1978) 349--362}.
  \url{https://www.sciencedirect.com/science/article/pii/0550321378903759}.

\bibitem{Carlip:2022pyh}
S.~Carlip, ``{Spacetime foam: a review},''  (9, 2022) ,
  \href{http://arxiv.org/abs/2209.14282}{{\ttfamily arXiv:2209.14282 [gr-qc]}}.

\bibitem{Planck:2013lks}
{\bfseries Planck} Collaboration, P.~A.~R. Ade {\em et~al.}, ``{Planck 2013
  results. XXIII. Isotropy and statistics of the CMB},''
  \href{http://dx.doi.org/10.1051/0004-6361/201321534}{{\em Astron. Astrophys.}
  {\bfseries 571} (2014) A23}, \href{http://arxiv.org/abs/1303.5083}{{\ttfamily
  arXiv:1303.5083 [astro-ph.CO]}}.

\bibitem{Planck:2015igc}
{\bfseries Planck} Collaboration, P.~A.~R. Ade {\em et~al.}, ``{Planck 2015
  results. XVI. Isotropy and statistics of the CMB},''
  \href{http://dx.doi.org/10.1051/0004-6361/201526681}{{\em Astron. Astrophys.}
  {\bfseries 594} (2016) A16},
  \href{http://arxiv.org/abs/1506.07135}{{\ttfamily arXiv:1506.07135
  [astro-ph.CO]}}.

\bibitem{Planck:2019evm}
{\bfseries Planck} Collaboration, Y.~Akrami {\em et~al.}, ``{Planck 2018
  results. VII. Isotropy and Statistics of the CMB},''
  \href{http://dx.doi.org/10.1051/0004-6361/201935201}{{\em Astron. Astrophys.}
  {\bfseries 641} (2020) A7}, \href{http://arxiv.org/abs/1906.02552}{{\ttfamily
  arXiv:1906.02552 [astro-ph.CO]}}.

\bibitem{Schwarz:2015cma}
D.~J. Schwarz, C.~J. Copi, D.~Huterer, and G.~D. Starkman, ``{CMB Anomalies
  after Planck},'' \href{http://dx.doi.org/10.1088/0264-9381/33/18/184001}{{\em
  Class. Quant. Grav.} {\bfseries 33} no.~18, (2016) 184001},
  \href{http://arxiv.org/abs/1510.07929}{{\ttfamily arXiv:1510.07929
  [astro-ph.CO]}}.

\bibitem{Abdalla:2022yfr}
E.~Abdalla {\em et~al.}, ``{Cosmology intertwined: A review of the particle
  physics, astrophysics, and cosmology associated with the cosmological
  tensions and anomalies},''
  \href{http://dx.doi.org/10.1016/j.jheap.2022.04.002}{{\em JHEAp} {\bfseries
  34} (2022) 49--211}, \href{http://arxiv.org/abs/2203.06142}{{\ttfamily
  arXiv:2203.06142 [astro-ph.CO]}}.

\bibitem{Sokolov:1974}
D.~D. Sokolov and V.~F. Shvartsman {\em Soviet Journal of Experimental and
  Theoretical Physics} {\bfseries 39} (1974) 196.

\bibitem{Fang:1983}
L.-Z. Fang and H.~Sato {\em Communications in Theoretical Physics,} {\bfseries
  2} (1983) 1055.

\bibitem{Fagundes:1987}
H.~V. Fagundes and U.~F. Wichoski {\em Nature,} {\bfseries 322} (1987) L5.

\bibitem{Lehoucq:1996qe}
R.~Lehoucq, M.~Lachieze-Rey, and J.~P. Luminet, ``{Cosmic crystallography},''
  {\em Astron. Astrophys.} {\bfseries 313} (1996) 339--346,
  \href{http://arxiv.org/abs/gr-qc/9604050}{{\ttfamily arXiv:gr-qc/9604050}}.

\bibitem{Roukema:1996cu}
B.~F. Roukema, ``{On determining the topology of the observable universe via
  3-d quasar positions},''
  \href{http://dx.doi.org/10.1093/mnras/283.4.1147}{{\em Mon. Not. Roy. Astron.
  Soc.} {\bfseries 283} (1996) 1147},
  \href{http://arxiv.org/abs/astro-ph/9603052}{{\ttfamily
  arXiv:astro-ph/9603052}}.

\bibitem{Weatherley:2003}
S.~J. Weatherley, S.~J. Warren, S.~M. Croom, {\em et~al.} {\em Nature,}
  {\bfseries 342} (2003) L9.

\bibitem{Fujii:2011ga}
H.~Fujii and Y.~Yoshii, ``{An improved cosmic crystallography method to detect
  holonomies in flat spaces},''
  \href{http://dx.doi.org/10.1051/0004-6361/201116521}{{\em Astron. Astrophys.}
  {\bfseries 529} (2011) A121},
  \href{http://arxiv.org/abs/1103.1466}{{\ttfamily arXiv:1103.1466
  [astro-ph.CO]}}.

\bibitem{Fujii:2013xsa}
H.~Fujii and Y.~Yoshii, ``{A search for nontoroidal topological lensing in the
  Sloan Digital Sky Survey quasar catalog},''
  \href{http://dx.doi.org/10.1088/0004-637X/773/2/152}{{\em Astrophys. J.}
  {\bfseries 773} (2013) 152}, \href{http://arxiv.org/abs/1306.2737}{{\ttfamily
  arXiv:1306.2737 [astro-ph.CO]}}.

\bibitem{Cornish:1996kv}
N.~J. Cornish, D.~N. Spergel, and G.~D. Starkman, ``{Circles in the Sky:
  Finding Topology with the Microwave Background Radiation},''  (2, 1996) ,
  \href{http://arxiv.org/abs/gr-qc/9602039}{{\ttfamily arXiv:gr-qc/9602039}}.

\bibitem{Cornish:1997ab}
N.~J. Cornish, D.~N. Spergel, and G.~D. Starkman, ``{Circles in the sky:
  Finding topology with the microwave background radiation},''
  \href{http://dx.doi.org/10.1088/0264-9381/15/9/013}{{\em Class. Quant. Grav.}
  {\bfseries 15} (1998) 2657--2670},
  \href{http://arxiv.org/abs/astro-ph/9801212}{{\ttfamily
  arXiv:astro-ph/9801212}}.

\bibitem{Cornish:1997hz}
N.~J. Cornish, D.~Spergel, and G.~Starkman, ``{Can COBE see the shape of the
  universe?},'' \href{http://dx.doi.org/10.1103/PhysRevD.57.5982}{{\em Phys.
  Rev. D} {\bfseries 57} (1998) 5982--5996},
  \href{http://arxiv.org/abs/astro-ph/9708225}{{\ttfamily
  arXiv:astro-ph/9708225}}.

\bibitem{Cornish:1997rp}
N.~J. Cornish, D.~N. Spergel, and G.~D. Starkman, ``{Measuring the topology of
  the universe},'' \href{http://dx.doi.org/10.1073/pnas.95.1.82}{{\em Proc.
  Nat. Acad. Sci.} {\bfseries 95} (1998) 82},
  \href{http://arxiv.org/abs/astro-ph/9708083}{{\ttfamily
  arXiv:astro-ph/9708083}}.

\bibitem{Riazuelo:2006tb}
A.~Riazuelo, S.~Caillerie, M.~Lachieze-Rey, R.~Lehoucq, and J.-P. Luminet,
  ``{Constraining cosmic topology with cmb polarization},''  (1, 2006) ,
  \href{http://arxiv.org/abs/astro-ph/0601433}{{\ttfamily
  arXiv:astro-ph/0601433}}.

\bibitem{deOliveira-Costa:2003utu}
A.~de~Oliveira-Costa, M.~Tegmark, M.~Zaldarriaga, and A.~Hamilton, ``{The
  Significance of the largest scale CMB fluctuations in WMAP},''
  \href{http://dx.doi.org/10.1103/PhysRevD.69.063516}{{\em Phys. Rev. D}
  {\bfseries 69} (2004) 063516},
  \href{http://arxiv.org/abs/astro-ph/0307282}{{\ttfamily
  arXiv:astro-ph/0307282}}.

\bibitem{Cornish:2003db}
N.~J. Cornish, D.~N. Spergel, G.~D. Starkman, and E.~Komatsu, ``{Constraining
  the topology of the universe},''
  \href{http://dx.doi.org/10.1103/PhysRevLett.92.201302}{{\em Phys. Rev. Lett.}
  {\bfseries 92} (2004) 201302},
  \href{http://arxiv.org/abs/astro-ph/0310233}{{\ttfamily
  arXiv:astro-ph/0310233}}.

\bibitem{ShapiroKey:2006hm}
J.~Shapiro~Key, N.~J. Cornish, D.~N. Spergel, and G.~D. Starkman, ``{Extending
  the WMAP Bound on the Size of the Universe},''
  \href{http://dx.doi.org/10.1103/PhysRevD.75.084034}{{\em Phys. Rev. D}
  {\bfseries 75} (2007) 084034},
  \href{http://arxiv.org/abs/astro-ph/0604616}{{\ttfamily
  arXiv:astro-ph/0604616}}.

\bibitem{Mota:2010jb}
B.~Mota, M.~J. Reboucas, and R.~Tavakol, ``{Circles-in-the-sky searches and
  observable cosmic topology in a flat Universe},''
  \href{http://dx.doi.org/10.1103/PhysRevD.81.103516}{{\em Phys. Rev. D}
  {\bfseries 81} (2010) 103516},
  \href{http://arxiv.org/abs/1002.0834}{{\ttfamily arXiv:1002.0834
  [astro-ph.CO]}}.

\bibitem{Bielewicz:2010bh}
P.~Bielewicz and A.~J. Banday, ``{Constraints on the topology of the Universe
  derived from the 7-year WMAP data},''
  \href{http://dx.doi.org/10.1111/j.1365-2966.2010.18057.x}{{\em Mon. Not. Roy.
  Astron. Soc.} {\bfseries 412} (2011) 2104},
  \href{http://arxiv.org/abs/1012.3549}{{\ttfamily arXiv:1012.3549
  [astro-ph.CO]}}.

\bibitem{Bielewicz:2011jz}
P.~Bielewicz, A.~J. Banday, and K.~M. Gorski, ``{Constraining the topology of
  the Universe using the polarised CMB maps},''
  \href{http://dx.doi.org/10.1111/j.1365-2966.2011.20371.x}{{\em Mon. Not. Roy.
  Astron. Soc.} {\bfseries 421} (2012) 1064},
  \href{http://arxiv.org/abs/1111.6046}{{\ttfamily arXiv:1111.6046
  [astro-ph.CO]}}.

\bibitem{Vaudrevange:2012da}
P.~M. Vaudrevange, G.~D. Starkman, N.~J. Cornish, and D.~N. Spergel,
  ``{Constraints on the Topology of the Universe: Extension to General
  Geometries},'' \href{http://dx.doi.org/10.1103/PhysRevD.86.083526}{{\em Phys.
  Rev. D} {\bfseries 86} (2012) 083526},
  \href{http://arxiv.org/abs/1206.2939}{{\ttfamily arXiv:1206.2939
  [astro-ph.CO]}}.

\bibitem{Aurich:2013fwa}
R.~Aurich and S.~Lustig, ``{A search for cosmic topology in the final WMAP
  data},'' \href{http://dx.doi.org/10.1093/mnras/stt924}{{\em Mon. Not. Roy.
  Astron. Soc.} {\bfseries 433} (2013) 2517},
  \href{http://arxiv.org/abs/1303.4226}{{\ttfamily arXiv:1303.4226
  [astro-ph.CO]}}.

\bibitem{Planck:2013kqc}
{\bfseries Planck} Collaboration, N.~Aghanim {\em et~al.}, ``{Planck 2013
  results. XXVII. Doppler boosting of the CMB: Eppur si muove},''
  \href{http://dx.doi.org/10.1051/0004-6361/201321556}{{\em Astron. Astrophys.}
  {\bfseries 571} (2014) A27}, \href{http://arxiv.org/abs/1303.5087}{{\ttfamily
  arXiv:1303.5087 [astro-ph.CO]}}.

\bibitem{Planck:2015gmu}
{\bfseries Planck} Collaboration, P.~A.~R. Ade {\em et~al.}, ``{Planck 2015
  results - XVIII. Background geometry and topology of the Universe},''
  \href{http://dx.doi.org/10.1051/0004-6361/201525829}{{\em Astron. Astrophys.}
  {\bfseries 594} (2016) A18},
  \href{http://arxiv.org/abs/1502.01593}{{\ttfamily arXiv:1502.01593
  [astro-ph.CO]}}.

\bibitem{Starkman_Priv_Comm}
P.~M. Vaudrevange, G.~D. Starkman, N.~J. Cornish, and D.~N. Spergel. {Private
  Communication}, 2013.

\bibitem{COMPACT:2022gbl}
{\bfseries COMPACT} Collaboration, Y.~Akrami {\em et~al.}, ``{The Promise of
  Future Searches for Cosmic Topology},''  (10, 2022) ,
  \href{http://arxiv.org/abs/2210.11426}{{\ttfamily arXiv:2210.11426
  [astro-ph.CO]}}.

\bibitem{COMPACT:2023rkp}
{\bfseries COMPACT} Collaboration, J.~R. Eskilt {\em et~al.}, ``{Cosmic
  topology. Part IIa. Eigenmodes, correlation matrices, and detectability of
  orientable Euclidean manifolds},''
  \href{http://arxiv.org/abs/2306.17112}{{\ttfamily arXiv:2306.17112
  [astro-ph.CO]}}.

\bibitem{Planck:2018vyg}
{\bfseries Planck} Collaboration, N.~Aghanim {\em et~al.}, ``{Planck 2018
  results. VI. Cosmological parameters},''
  \href{http://dx.doi.org/10.1051/0004-6361/201833910}{{\em Astron. Astrophys.}
  {\bfseries 641} (2020) A6}, \href{http://arxiv.org/abs/1807.06209}{{\ttfamily
  arXiv:1807.06209 [astro-ph.CO]}}. [Erratum: Astron. Astrophys. 652, C4
  (2021)].

\bibitem{Hawking:1973uf}
S.~W. Hawking and G.~F.~R. Ellis,
  \href{http://dx.doi.org/10.1017/CBO9780511524646}{{\em {The Large Scale
  Structure of Space-Time}}}.
\newblock Cambridge Monographs on Mathematical Physics. Cambridge University
  Press, 2, 2011.

\bibitem{arthive}
``Flying fish (no. 73), 1949 by maurits cornelis escher: History, analysis \&
  facts.''
\newblock \url{https://arthive.com/escher/works/200144~Flying_Fish_No_73}.

\bibitem{Riazuelo:2003ud}
A.~Riazuelo, J.~Weeks, J.-P. Uzan, R.~Lehoucq, and J.-P. Luminet, ``{Cosmic
  microwave background anisotropies in multi-connected flat spaces},''
  \href{http://dx.doi.org/10.1103/PhysRevD.69.103518}{{\em Phys. Rev. D}
  {\bfseries 69} (2004) 103518},
  \href{http://arxiv.org/abs/astro-ph/0311314}{{\ttfamily
  arXiv:astro-ph/0311314}}.

\bibitem{Thurston1982ThreeDM}
W.~P. Thurston, ``Three dimensional manifolds, kleinian groups and hyperbolic
  geometry,'' {\em Bulletin of the American Mathematical Society} {\bfseries 6}
  (1982) 357--381.

\end{thebibliography}\endgroup
\end{document}